\documentclass{jltp}
\usepackage{graphicx}
\runninghead{W. J. Mullin, M. Holzmann, and F. 
Lalo\"{e}}{Instability in a Two-Dimensional Dilute Interacting Bose System}

\title{Instability in a Two-Dimensional Dilute Interacting Bose System}
\author{W. J. Mullin$^{a}$, M. Holzmann$^{b}$, and F. Lalo\"{e}$^{b}$ \\}
\address{$^{a}$Physics Department, Hasbrouck Laboratory\\
University of Massachusetts Amherst, MA 01003 USA\\
$^{b}$LKB, D\'{e}partement de Physique de l'ENS\\
24 rue Lhomond 75005 Paris, France\\
}
\begin{document}
\maketitle

\begin{abstract}
\noindent The formalism of Ursell operators provides a self-consistent
integral equation for the one-particle reduced operator.  In three
dimensions this technique yields values of the shift in the
Bose-Einstein condensation (BEC) transition temperature, as a function
of the scattering length, that are in good agreement with those of
Green's function and quantum Monte Carlo methods.  We have applied the
same equations to a uniform two-dimensional system and find that, as we
alter the chemical potential, an instability develops so that the
self-consistent equations no longer have a solution.  This
instability, which seems to indicate that interactions restore a
transition, occurs at a non-zero value of an effective chemical
potential.  The non-linear equations are limited to temperatures
greater than or equal to Tc, so that they do not indicate the nature
of the new stable state, but we speculate concerning whether it is a
Kosterlitz-Thouless state or a ``smeared'' BEC, which might avoid any
violation of the Hohenberg theorem, as described in an accompanying
paper.

\end{abstract}

\section{INTRODUCTION}

Interest in Bose systems and Bose-Einstein condensation (BEC) has been
expanding in recent years, stimulated in part by the experimental work
on alkali atoms in magnetic traps.\cite{Alkali} Much of the
theoretical work has involved mean-field theories, such as those based
on the Gross-Pitaevskii equation.\cite{Stringari} However, some
theoretical analyses have recently reconsidered the properties of
three-dimensional (3D) homogeneous systems with the inclusion of
dynamic correlations that go beyond mean-field theory. Here we extend
the methods to dilute systems in two-dimensions (2D).

In 3D, these techniques have provided computations of the change in
the BEC transition temperature as a function of the repulsive
interactions in a dilute Bose
gas.\cite{HGL}$^{,}$\cite{BBHLV}$^{,}$\cite{BahmBlZin} Ref.\
 %online 
\cite{HGL} reports results from an Ursell operator
approach,\cite{Ursell123} which are confirmed by non-perturbative
Green's function methods.\cite{BBHLV}$^{,}$\cite{BahmBlZin} The two
techniques give equivalent non-linear self-consistent equations for
the self-energies and the distribution functions.  Each method finds
that a dilute Bose system with repulsive interactions reduces its
self-energy by spatial rearrangements of particles with low
velocities, and that the low-$k$ quasi-particle energies behave as
$k^{2-\eta}$ near the transition.  (The constant $\eta$ is equal to
$1/2$ in lowest order approximation, and smaller in higher order.) 
Both methods predict that there is an increase of the critical
temperature as the repulsive interactions are turned on, a result
confirmed by quantum Monte Carlo calculations.\cite{HoltzKrauth}

There may soon be a number of experiments involving 2D
systems for, say, alkali atoms or for spin-polarized hydrogen in traps
or in homogeneous settings.  The usefulness of the Ursell approach in
3D suggests studying 2D systems by the same approach.  Here we
solve the self-consistent equation for the self-energy as a function
of the chemical potential.  In 3D, as the chemical potential is
increased from large negative values, one reaches a critical value
where the phase transition takes place; an effective
chemical potential that includes the effects of mean-field and
dynamic correlations vanishes signaling the transition.  We carry out
a similar program here and find that at a finite value of the
effective chemical potential the equations no longer have a
self-consistent solution, again signaling some sort of transition.

The Hohenberg theorem\cite{Hohenberg} states that a 2D system cannot
undergo a BEC into a single state, but recent work shows that a
``smeared'' condensation is not forbidden by this theorem.\cite{MHL}
In this mode of BEC, no single state is extensively occupied, but each
of the states in a band has a large, although non-extensive, occupation,
with enough states in the band that its total occupation is extensive. 
Furthermore, one knows that superfluidity occurs in 2D via a
Kosterlitz-Thouless transition.\cite{KT} It is even possible that the
KT transition and a smeared BEC are equivalent.\cite{Popov} Since our
theory has not yet been extended to regimes in which there is
macroscopic occupation of a single state or of a band of states, we are
unable to tell what transition is signaled by our instability.

\section{URSELL OPERATOR APPROACH }

We use the method of Ursell operators,\cite{Ursell123} an approach
that allows the generalization of the usual cluster expansion to all
orders in statistical effects, while still allowing the possiblity of
including interactions in lower order.  A basic
quantity is the second Ursell operator given by
\begin{equation}
U_{2}(1,2)=e^{-\beta H_{2}(1,2)}-e^{-\beta (H_{1}(1)+H_{1}(2))}
\end{equation}
where $H_{1}$ is a single-particle Hamiltonian and $H_{2}$ a
two-particle Hamiltonian.  A particular problem in 2D is that the matrix
elements of this operator do not approach a constant at small $k$;
indeed, for hard-core scattering, the leading term is proportional to
$1/\ln (ka)$, where $a$ is the hard-core radius.  However, we will see
that we can, to good approximation, avoid that complication and use
a k-independent $U_{2}$ matrix element, as in 3D.

The Ursell operator method gives an integral equation for the
one-particle reduced density $\rho _{1}$ with Fourier transform $\rho
_{k}$.  When the occupation of all states is small (no BEC), one
finds,\cite{HGL} to lowest order in the interaction, that $\rho _{k}$
is given by a free-particle distribution function with a
self-consistent, $k$-dependent effective chemical potential:
$%\begin{equation}
\rho _{k}=\left[ e^{\beta (\varepsilon _{k}-\tilde{\mu}_{k})}-1\right] 
^{-1},
$ %\end{equation}
where $\varepsilon _{k}=\hbar ^{2}k^{2}/2m$ and $\tilde{\mu}_{k}=\mu
-\Delta \mu _{k}$ is the effective chemical potential. The leading
 correction to the chemical potential is given by $\beta \Delta
\mu _{k}=-\ln \left[ 1-\Delta \xi _{k}\right] ,$ where 
\begin{equation}
\Delta \xi _{k}=-\frac{2\lambda ^{2}}{(2\pi )^{2}}\int
d\mathbf{k}^{\prime}\frac{\rho _{k}e^{\beta \Delta \mu
_{k^{\prime}}}}{\ln \left(\left| \mathbf{k}-\mathbf{k}^{\prime
}\right| \bar{a}\right)}
\end{equation}
with $\bar{a}$ equal to $a$ times a constant of order unity.  (In 3D
the log factor does not occur and so $\Delta \xi $ is independent of
$k.$) We have solved these equations by iteration and find that
$\Delta \xi _{k}$ is a very weak function of $k,$ so that one might as
well set the logarithm to a constant.  A constant correction to the
chemical potential or, equivalently, the single-particle energy is
simply a mean-field, which can cause no fundamental change, and
certainly no phase transition, in the Bose gas.  Ignoring the
logarithm is equivalent to using a momemtum-independent matrix element
of $U_{2}$, which is what we use in the next order approximation.

If we add corrections second order in the interaction we
necessarily develop velocity-dependent terms in the distribution
function.  We find that the effective chemical potential change is
\begin{equation}
\beta \Delta \mu _{k}=-\ln \left[ 1-\Delta \xi _{o}-\delta \xi _{k}\right] 
\label{chempot}
\end{equation}
where, with our assumption of constant interaction, we now have 
\begin{equation}
\Delta \xi _{0}=\frac{\alpha}{(2\pi )^{2}}\int d\mathbf{k}\rho _{%
\mathbf{k}}e^{\beta \Delta \mu _{k}}
\end{equation}
with $\alpha $ a parameter measuring the strength of the interaction.
The second-order correction in Eq.(\ref{chempot}) is 
\begin{equation}
\delta \xi _{\mathbf{k}}=-\frac{\alpha ^{2}}{2(2\pi )^{4}}\int d\mathbf{k}%
^{\prime }\rho _{\mathbf{k}^{\prime }}e^{\beta \Delta \mu _{\mathbf{k}%
^{\prime }}}\int d\mathbf{q}\rho _{\mathbf{k}^{\prime }-\mathbf{q}}\rho _{%
\mathbf{k}+\mathbf{q}}e^{\beta (\Delta \mu _{\mathbf{k}^{\prime }-\mathbf{q}%
}+\Delta \mu _{\mathbf{k}+\mathbf{q}})}  \label{delxi}
\end{equation}
The quantities $\Delta \xi _{0}$ and $\delta \xi _{\mathbf{k}}$ are
equivalent to self-energies in the Green's function approach.\cite{BBHLV}

There are a variety of calculational procedures that can be followed
to compute $\delta \xi _{\mathbf{k}}$.  If one assumes that $\rho
_{\mathbf{k}}$ is dependent only on the magnitude of $\mathbf{k}$,
then this quantity can be reduced to a four-fold numerical integral. 
By changes of variables this integral can be done in several ways. 
Alternatively, one can introduce the Fourier transform of $%
\rho _{\mathbf{k}}e^{\beta \Delta \mu _{k}}$; since Eq.(\ref{delxi})
is a convolution, the computation reduces to doing the Fourier
transform and a one-dimensional integral.  The procedure used is to
pick a value of the real chemical potential, compute the interaction
corrections, and iterate.  As in the 3D case, $\delta \xi
_{\mathbf{k}}$ has a minimum at $k=0$ and rises monotonically.  As one
increases $\mu $ from very negative values, as would happen in 3D in a
system approaching criticality, the minimum in $%
\delta \xi _{\mathbf{k}}$ deepens and the curvature at $k=0$
sharpens.  The value of $\left| \tilde{\mu}_{k}\right| $ decreases
during this process.  This feature is caused by dynamic correlations
that lead to rearrangements of particles with small $k.$

Unexpectedly, at a particular value of $\mu $, with $\left| \tilde{\mu}%
_{k}\right| $ still \emph{nonvanishing}, the iteration becomes
unstable and the self-consistent solution disappears as shown in Fig.\
1.  This result is quite robust and occurs independently of the
numerical method used.  At the instability $|\delta \xi _{0}|$,
instead of converging to a self-consistent value as one iterates,
increases rapidly until the process predicts a positive value of
$\tilde{\mu}_{0}$, at which point the iterations must halt.  Passing 
through
$\tilde{\mu}_{0}=0$ would seem to indicate the onset of BEC. Since the
theory has not been extended into any condensed regime we are unable
to say what the new stable phase might be.

%\begin{figure}[h]
%\centerline{\includegraphics[height=2.5in]{Deltamu.epsf}} 
%framebox[5in]{%
%\rule[1.125in]{0in}{1.125in}} makebox[5in]{%
%\rule[1.125in]{0in}{1.125in}}
%\label{fig:Deltamu}
%\end{figure}

Since one knows that there is a KT phase in 2D, one might expect that our
theory is showing an instability towards that phase. A second possibility
arises from the proof in an accompanying paper,\cite{MHL} which shows
that the Hohenberg theorem does not rule out what we term as a ``smeared''
BEC, as defined in the Introduction. It will be
interesting to extend the theory to cover such a possibility. 

One can go to higher orders in the approximation scheme and sum all
``bubble'' diagrams, in which case one finds
\begin{equation}
\delta \xi _{\mathbf{k}}=\frac{\alpha ^{2}}{2(2\pi )^{4}}\int d\mathbf{k}%
^{\prime }\rho _{\mathbf{k}^{\prime }}e^{\beta \Delta \mu _{\mathbf{k}%
^{\prime }}}\left\{ \frac{\int d\mathbf{q}\rho _{\mathbf{q}}\rho _{\mathbf{k}%
+\mathbf{q}+\mathbf{k}^{\prime }}e^{\beta (\Delta \mu _{\mathbf{q}}+\Delta %
\mu _{\mathbf{k}+\mathbf{q}+\mathbf{k}^{\prime }})}}{1-\frac{\alpha ^{2}}{2}%
\int d\mathbf{q}\rho _{\mathbf{q}}\rho _{\mathbf{k}+\mathbf{q}+\mathbf{k}%
^{\prime }}e^{\beta (\Delta \mu _{\mathbf{q}}+\Delta \mu _{\mathbf{k}+%
\mathbf{q}+\mathbf{k}^{\prime }})}}\right\} 
\end{equation}
We find that the instability persists, although the interations now have a
somewhat different behavior. As one iterates, values of $\Delta \xi _{0}$
and $\delta \xi _{\mathbf{k}}$ conspire to make the subsequent values of
these quantities oscillate about an average value. Either this oscillation
dies out eventually to a self-consistent value or grows until it becomes
completely unstable.  

\section{DISCUSSION}

Because of the success of the Ursell approach in 3D, namely, its good
agreement with the results of other powerful approaches, we have
confidence that it should yield reliable results in 2D. Our results
show an instability in this case, but we are unable, at this time, to
determine if this is the signal of a true phase transition, what is
the order of such a transition, or what might be the stable state to
which the system proceeds.  It is known that perturbation theory
breaks down near the BEC phase
transition\cite{BBHLV}$^{-}$\cite{Ursell123}, and that
applies here as well.  Nevertheless, the unknown sum of the diagrams
associated with interparticle correlations must give some negative
result, because the mean-field theory can be seen as a variational
evaluation with the same wave function as an ideal gas, and the
inclusion of correlations can only reduce the free energy and bring
the system closer to intability.  As $\mu$ and the density increase, it
seems then that an instability must ultimately take place. 
Possiblities are that we are seeing a KT transition or, perhaps more
interestingly, a transition to a ``smeared'' Bose-Einstein
condensation, which our accompanying article shows is
possible.\cite{MHL} Thus the present paper is really a progress report
and we will need further research to resolve the questions it raises.

\section{ACKNOWLEDGMENTS}
The Laboratoire Kastler-Brossel is Unit\'{e} Associ\'{e}e au CNRS
(UMR8552) et \`{a} l'Universit\'{e} Pierre et Marie Curie.  WJM would like
to thank Ecole Normale Sup\'{e}rieure, where some of this research was
carried out, for a travel grant and for its excellent hospitality.

\section{FIGURE CAPTION}
FIG. 1 Change in the effective chemical potential for $k=0$ as a
function of the real chemical potential.  For the interaction strength
used the instability occurs at $\beta\mu>0.13$, beyond which the plot
shows successive final iteration values just before failure rather
than self-consistent values.  BEC criticality occurs when $\mu+\Delta
\mu=0$, i.e., when the $\Delta \mu_{0}$ touches the diagonal $\mu$ line.


\begin{thebibliography}{9}
\bibitem{Alkali} M. H. Anderson, J. R. Ensher, M. R. Matthews, C. E.
Wieman, and E. A. Cornell, {\it Science} {\bf 269}, 198 (1995); K. B.
Davis, M.-O. Mewes, M. R. Andrews, N. J. van Druten, D. S. Durfee, D.
M. Kurn, and W. Ketterle, {\it Phys.  Rev. Lett.} {\bf 75}, 3969 
(1995); C. C.
Bradley, C. A. Sackett, J. J. Tollett, and R. G. Hulet, {\it Phys.  
Rev. Lett.} {\bf 75}, 1687 (1995).

\bibitem{Stringari} F. Dalfovo, S. Giorgini, L. Pitaevskii, and S. Stringari, 
\textit{Rev. Mod. Phys.} \textbf{71}, 463 (1999). 

\bibitem{HGL} M. Holzmann, P. Gr\"{u}ter, and F. Lalo\"{e},\textit{
Europhys. Journ. B} \textbf{10}, 739 (1999).

\bibitem{BBHLV} G. Baym, J.-P. Blaizot, M. Holzmann, F. Lalo\"{e}, 
and D. Vautherin, \textit{Phys. Rev. Lett.} \textbf{83}, 1703 (1999).

\bibitem{BahmBlZin} G. Baym, J.-P. Blaizot, 
and J. Zinn-Justin, \textit{Europhys. Lett.} \textbf{49}, 150 (2000).

\bibitem{Ursell123}P. Gr\"{u}ter and F. Lalo\"{e}, {\it J. Phys. 
I France} \textbf{5}, 181 (1995); \textbf{5}, 1255 (1995); P.
Gr\"{u}ter, F. Lalo\"{e}, A. E. Meyerovich, and W.
Mullin, {\it J. Phys.  I France} \textbf{7}, 485 (1997); F. Lalo\"{e}, 
in {\it Bose-Einstein Condensation}, edited by A. Griffin, D. W. Snoke, 
and S. Stringari, Cambridge Univ. Press (1995).

\bibitem{HoltzKrauth} M. Holzmann and W. Krauth, \textit{Phys.  Rev. 
Lett.} \textbf{83}, 2687 (1999).

\bibitem{KT}  J. M. Kosterlitz and D. J. Thouless, \textit{J. Phys. C }%
\textbf{6}, 1181 (1973); V. L. Berezinskii, \textit{Sov. Phys. JETP \textbf{
32, }}493 (1971).

\bibitem{Hohenberg}P.C. Hohenberg,\textit{\ Phys. Rev.} \textbf{158}, 383
(1967).

\bibitem{MHL}W. J. Mullin,  M. Holzmann, and F. Lalo\"{e}, (This 
conference).

\bibitem{Popov}  V. N. Popov, \textit{Functional Integrals in Quantum Field
Theory and Statistical Physics,} (Reidel Publishing Company, Dordrecht,
1983).



\end{thebibliography}
\end{document}